\pdfoutput=1

\documentclass[11pt]{article}

\usepackage{adjustbox} 
\usepackage{amsmath,amssymb,amsfonts}
\usepackage{algorithm}        
\usepackage{algpseudocode}    
\usepackage{graphicx}
\usepackage{hyperref}
\usepackage{enumitem}
\usepackage{tabularx} 

\usepackage{booktabs}
\usepackage{amsmath, amssymb}
\usepackage{multirow}
\usepackage{textcomp}

\usepackage{algorithm}
\usepackage{algpseudocode}

\usepackage[preprint]{ACL}

\usepackage{times}
\usepackage{latexsym}

\usepackage[T1]{fontenc}

\usepackage[skins]{tcolorbox}

\usepackage[utf8]{inputenc}

\usepackage{microtype}

\usepackage{inconsolata}

\usepackage{graphicx}
\def\code#1{\texttt{#1}}
\title{An Empirical Study of Position Bias in Modern Information Retrieval}

\author{Ziyang Zeng\textsuperscript{\rm 1,2} \quad Dun Zhang\textsuperscript{\rm 2,3}\thanks{~~Project leader.}  \quad Jiacheng Li\textsuperscript{\rm 2} \\ \textbf{Panxiang Zou}\textsuperscript{\rm 4} \quad
\textbf{Yudong Zhou}\textsuperscript{\rm 3} 
\quad \textbf{Yuqing Yang}\textsuperscript{\rm 1}\thanks{~~Corresponding author.}
\\
\textsuperscript{\rm 1}Beijing University of Posts and Telecommunications \\  
\textsuperscript{\rm 2}NovaSearch Team \quad \textsuperscript{\rm 3}Prior Shape \quad \textsuperscript{\rm 4}RichInfo\\ 
\texttt{ziyang1060@bupt.edu.cn, \{dunnzhang0,jcli.nlp\}@gmail.com} \\ 
\texttt{zoupanxiang@richinfo.cn, zhouyudong@priorshape.com} \\
\texttt{yangyuqing@bupt.edu.cn}
}

\begin{document}

\maketitle

\begin{abstract}
This study investigates the \textit{position bias} in information retrieval, where models tend to overemphasize content at the beginning of passages while neglecting semantically relevant information that appears later. 
To analyze the extent and impact of position bias, we introduce a new evaluation framework consisting of two position-aware retrieval benchmarks (\textsc{SQuAD-PosQ}, \textsc{FineWeb-PosQ}) and an intuitive diagnostic metric, the Position Sensitivity Index (PSI), for quantifying position bias from a worst-case perspective.
We conduct a comprehensive evaluation across the full retrieval pipeline, including BM25, dense embedding models, ColBERT-style late-interaction models, and full-interaction reranker models. 
Our experiments show that when relevant information appears later in the passage, dense embedding models and ColBERT-style models suffer significant performance degradation (an average drop of 15.6\%).
In contrast, BM25 and reranker models demonstrate greater robustness to such positional variation.
These findings provide practical insights into model sensitivity to the position of relevant information and offer guidance for building more position-robust retrieval systems.
Code and data are publicly available at: \url{https://github.com/NovaSearch-Team/position-bias-in-IR}.
\end{abstract}

\section{Introduction}

\begin{figure}
 \centering
\includegraphics[width=1\columnwidth]{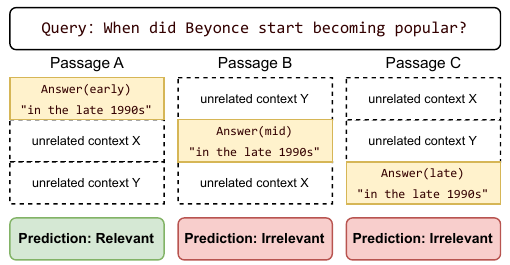} 
\caption{Illustration of position bias in IR: models often focus disproportionately on the beginning of passages, overlooking relevant content that appears later.}
\label{fig:illustration}
\end{figure}

Information Retrieval (IR) underpins a broad range of applications, such as web search~\cite{croft2010search}, question answering~\cite{search_qa}, and Retrieval-Augmented Generation (RAG)~\cite{RAG1}. 
A central challenge for IR systems is to accurately assess the semantic relevance between user queries and candidate passages. 
Recent advances in neural IR models, particularly those leveraging pre-trained language models such as BERT~\cite{bert}, have significantly improved retrieval performance~\cite{BERTandBeyond,10.1007/978-3-030-72240-1_26}. 
However, prior work~\cite{MitigatingthePositionBias,HowDoesBERTRerankPassages,ABNIRML} has identified that such models exhibit a \emph{position bias}: a tendency to overemphasize content at the beginning of passages, while overlooking semantically relevant information appearing later.
Figure~\ref{fig:illustration} illustrates how such position bias can lead retrieval models to systematically underestimate the relevance of passages, especially when key information appears later, which may harm downstream performance (e.g., in RAG~\cite{CollapseEmbedding}) or expose vulnerabilities to adversarial attacks~\cite{attack2022}.

While modern neural IR models have seen significant advances in architecture and training techniques~\cite{RepLlama,lee2025nvembed}, recent studies suggest that position bias remains present in a few specific embedding models~\cite{coelho-etal-2024-dwell,CollapseEmbedding}.  
This motivates a broader research question: \textit{How prevalent is position bias among today’s state-of-the-art IR models, and how does this bias manifest across different IR architectures?}
Answering this question requires a systematic investigation across diverse classes of retrieval models—an area that remains largely underexplored.
Meanwhile, the findings from recent studies on position bias are often based on synthetically manipulated passages—e.g., by inserting relevant spans at predetermined positions~\cite{coelho-etal-2024-dwell,CollapseEmbedding}. 
While these controlled setups are diagnostically useful, they risk introducing semantic discontinuities and may not reflect realistic retrieval conditions.

To this end, we introduce two new position-aware English retrieval benchmarks—\textsc{SQuAD-PosQ} and \textsc{FineWeb-PosQ}—designed to evaluate model performance in retrieval scenarios where relevant content appears at varying positions within passages.
Derived from SQuAD v2~\cite{rajpurkar-etal-2018-know} and FineWeb-edu~\cite{FineWeb}, respectively, these benchmarks differ in passage length and construction methodology, offering complementary perspectives for analysis.
We preserve the original passage structure and construct position-sensitive questions either from existing annotated QA pairs or by prompting large language models (LLMs)~\cite{llm_survey} to target specific passage regions.
A two-stage filtering pipeline is applied to validate positional relevance and minimize false negatives~\cite{AIRBench} during question generation.
To accompany these benchmarks with a quantifiable diagnostic, we propose a simple and intuitive metric, the Position Sensitivity Index (PSI), which provides a worst-case perspective by explicitly quantifying the maximum relative degradation across positions.

We perform a comprehensive evaluation across diverse IR models to analyze the extent and impact of position bias, including sparse retrievers (e.g., BM25), dense embedding-based retrievers, ColBERT-style late-interaction models, and full-interaction reranker models. 
Our results reveal that dense embedding and ColBERT-style models exhibit an average 15.6\% performance drop when relevant content is located later in the passage, revealing a consistent bias toward early-passage content.
In contrast, BM25 and reranker models remain largely robust to positional shifts.
These differences highlight architectural sensitivities to where relevant information appears in a passage.


\section{Related Work}
Prior work has extensively documented the existence of position bias in neural IR systems.
\citet{MitigatingthePositionBias} observe that in MS MARCO~\cite{MSMARCO}, a widely used collection in the IR community, answer spans are disproportionately concentrated in the earlier portions of passages.
BERT-based neural IR models fine-tuned on MS MARCO tend to inherit and reinforce this position bias~\cite{HowDoesBERTRerankPassages,ABNIRML}, potentially leading to overestimated performance due to shared distributional artifacts between training and evaluation sets~\cite{RevisitingBagofWords}.
\citet{coelho-etal-2024-dwell} further dissect the training pipeline of a T5-based dense retriever on MS MARCO, showing that position bias primarily originates during contrastive pre-training and is amplified in contrastive fine-tuning. 
\citet{CollapseEmbedding} extend this analysis by uncovering multiple forms of bias—including position bias—in dense embedding models, and linking them to generation failures in RAG pipelines.
Beyond IR, similar position-sensitive behaviors have been observed in LLMs, particularly in how attention is distributed across long contexts~\cite{liu-etal-2024-lost,10.5555/3737916.3739859,wu2025on}, suggesting that position bias may be a more general limitation of modern transformer architectures.

\section{Position-Aware Retrieval Benchmarks}

\subsection{Repurposing Existing QA Pairs}
We repurpose the Stanford Question Answering Dataset v2 (SQuAD v2), leveraging its character-level answer span annotations for fine-grained positional analysis. 
After removing unanswerable questions—originally designed to probe abstention behaviors—we obtain 92,749 examples, each represented as a (\textit{question}, \textit{passage}, \textit{answer\_start\_position}\footnote{Refers to the index position of the first character of the answer, with the length calculated in units of characters.}) triple. 
We denote this dataset as \textsc{SQuAD-PosQ}.
To analyze position bias, we bucket the questions into six groups based on the character-level start index of their answers: [0–100], [100–200], [200–300], [300–400], [400–500], and [500–3120]\footnote{The maximum observed index is 3120.}, where all intervals are inclusive. 
We choose fixed-width intervals of 100 characters to enable fine-grained comparison of retrieval performance across different passage regions, while ensuring sufficient examples in each bucket.
Each question is treated as a query, with its gold passage designated as the relevant target in a passage ranking task over the full retrieval corpus.
This setup enables us to assess how the position of relevant content affects retrieval accuracy under realistic conditions.
A consistent performance drop for questions where the answer appears later in the passage would indicate the presence of position bias.
For efficiency,, we additionally construct a smaller subset, \textsc{SQuAD-PosQ-Tiny}, consisting of 10,000 randomly sampled triplets, while keeping the retrieval corpus unchanged.

\begin{table*}[!t]
  \centering
  \begin{adjustbox}{max width=\textwidth}
  \begin{tabular}{lccccccc|cccc}
    \toprule
    \multirow{2}{*}{\textbf{Retrieval Models}} 
      & \multicolumn{7}{c}{\textbf{SQuAD-PosQ }} 
      & \multicolumn{4}{c}{\textbf{FineWeb-PosQ }} \\
    \cmidrule(lr){2-8} \cmidrule(lr){9-12}
      & 0+    & 100+    & 200+   & 300+    & 400+    & 500+ & \textit{PSI $\downarrow$} & begin & middle & end  & \textit{PSI $\downarrow$}\\
    \midrule
    \multicolumn{10}{l}{\textbf{Sparse Retrievers}} \\
    BM25 
      & 76.62 & 79.37 & 80.61 & 81.06 & 81.43 & 79.49 & 0.059
      & 89.40  & 90.80  & 88.36 & 0.027 \\
    \addlinespace
    \multicolumn{10}{l}{\textbf{Dense Embedding-based Retrievers}} \\
    bge-m3-dense\textsuperscript{*}
      & 84.47 & 83.03 & 81.47 & 79.95 & 77.98 & 74.61 & 0.117
      & 88.77  & 78.39  & 71.88 & 0.190 \\
    stella\_en\_400M\_v5\textsuperscript{*}
      & 85.78 & 83.62 & 82.24 & 80.34 & 78.96 & 75.69 & 0.118
      & 86.10  & 77.92  & 69.41 & 0.194 \\
    Qwen3-Embedding-0.6B\textsuperscript{*}
      & 82.60 & 81.93 & 79.08 & 77.36 & 75.39 & 71.48 & 0.135
      & 88.54  & 78.83  & 65.61 & 0.259 \\
    text-embedding-3-large\textsuperscript{*}
      & 85.19 & 82.45 & 80.32 & 77.84 & 75.27 & 71.10 & 0.165
      & 81.72  & 75.95  & 79.50 & 0.071 \\
     voyage-3-large\textsuperscript{*}
      & 89.93 & 89.32 & 89.17 & 88.70 & 88.09 & 86.73 & 0.036
      & 92.76  & 87.46  & 83.38 & 0.101 \\
    jina-embeddings-v4\textsuperscript{*}
      & 82.50 & 80.55 & 78.87 & 77.33 & 75.91 & 72.94 & 0.116
      & 88.35  & 77.46  & 69.80 & 0.210 \\
    Qwen3-Embedding-4B\textsuperscript{*}
      & 86.36 & 85.92 & 85.17 & 83.77 & 82.09 & 78.85 & 0.087
      & 89.74  & 81.48  & 70.72 & 0.212 \\
    gte-Qwen2-7B-instruct\textsuperscript{*}
      & 85.13 & 83.85 & 83.33 & 81.71 & 80.13 & 77.75 & 0.087
      & 84.24  & 79.07  & 75.90 & 0.099 \\
    NV-embed-v2\textsuperscript{*}
      & 93.04 & 93.55 & 93.48 & 93.02 & 92.48 & 90.72 & 0.030
      & 77.24  & 85.12  & 85.98 & 0.102 \\
    Qwen3-Embedding-8B\textsuperscript{*}
      & 89.16 & 87.55 & 85.90 & 84.05 & 82.13 & 78.82 & 0.116
      & 90.80  & 83.35  & 73.66 & 0.189 \\
    \addlinespace
    \multicolumn{10}{l}{\textbf{ColBERT-style Late-interaction Models}} \\
    colbertv2.0\textsuperscript{*}
      & 91.85 & 90.27 & 91.74 & 89.64 & 86.71 & 84.57 & 0.079
      & -  & -  & - & - \\
    jina-colbert-v2\textsuperscript{*}
      & 93.52 & 92.42 & 93.28 & 92.58 & 91.80 & 78.14 & 0.164
      & 91.69  & 56.45  & 45.91 & 0.499 \\
    bge-m3-colbert\textsuperscript{*} 
      & 89.88 & 88.09 & 88.84 & 87.68 & 86.72 & 86.36 & 0.039
      & 92.77  & 86.38  & 81.82 & 0.118 \\
    \addlinespace
    \multicolumn{10}{l}{\textbf{Full-interaction Reranker Models}} \\
    bge-reranker-v2-m3 
      & 93.53 & 93.56 & 94.69 & 94.50 & 94.42 & 94.52 & 0.012
      & 94.25  & 96.10  & 94.87 & 0.019 \\
    Qwen3-Reranker-0.6B
      & 92.11 & 91.43 & 91.53 & 91.65 & 90.60 & 89.69 & 0.026
      & 95.03  & 94.97  & 92.46 & 0.027 \\
    gte-multilingual-reranker
      & 90.70 & 91.10 & 92.59 & 91.84 & 91.57 & 92.03 & 0.020
      & 94.70  & 95.73  & 95.51 & 0.011 \\
    Qwen3-Reranker-4B
      & 93.32 & 92.84 & 93.38 & 93.94 & 92.57 & 93.26 & 0.015
      & 95.06  & 96.58  & 95.23 & 0.016 \\
    bge-reranker-v2-gemma
      & 94.31 & 94.01 & 94.73 & 94.80 & 94.55 & 94.55 & 0.008
      & 94.38  & 95.84  & 96.02 & 0.017 \\
    Qwen3-Reranker-8B
      & 93.38 & 93.48 & 93.81 & 94.20 & 93.83 & 94.31 & 0.010
      & 95.61  & 97.02  & 96.74 & 0.015 \\
    \bottomrule
  \end{tabular}
  \end{adjustbox}
  \caption{NDCG@10 scores~$\uparrow$ and Position Sensitivity Index (PSI)~$\downarrow$ of retrieval models on \textsc{SQuAD-PosQ} and \textsc{FineWeb-PosQ}. Models exhibiting notable position bias (i.e., PSI~$\ge 0.03$ on both datasets) are marked with \textsuperscript{*}.}
  \label{tab:retrieval-results}
\end{table*}

\subsection{Generating Position-sensitive Questions}
While \textsc{SQuAD-PosQ} serves as a useful benchmark to analyze position bias, it has two key limitations: (1) its passages are relatively short (averaging 117 words), and (2) it is likely included in the training data of many retrieval models~\cite{bge-m3,lee2025nvembed}, raising concerns about evaluation leakage.  
To address these issues, we construct a synthetic dataset using passages from the FineWeb-edu, a large-scale, high-quality educational web text corpus.
We sample 13,902 passages from the collection whose lengths range from 500 to 1,024 words.
We instruct \texttt{gpt-4o-mini}~\cite{4omini} to generate questions anchored to localized chunks of each passage, following carefully designed prompts (see Appendix~\ref{sec:prompts}).
Each passage is divided into three equal-length segments—\textit{beginning}, \textit{middle}, and \textit{end}—and each question is assigned to one of these buckets based on the position of its supporting chunk.
Additionally, we apply a two-stage filtering pipeline to ensure high-quality question generation by validating positional relevance and minimizing false negatives (see Appendix~\ref{sec:appendix_fineweb_construction}).
The resulting dataset, \textsc{FineWeb-PosQ}, contains 25,775 synthetic questions and facilitates rigorous evaluation of position sensitivity in longer-context retrieval.
For efficiency, we also create a smaller version, \textsc{FineWeb-PosQ-Tiny}, by sampling 1,000 questions from each position category, resulting in a total of 3,000 questions. 

Appendix~\ref{sec:data_info} provides more details on dataset statistics, construction methodology, and empirical validation of the sampled subsets as reliable proxies for evaluating position bias.

\subsection{Position Sensitivity Index}
\label{sec:psi}
To quantify a retrieval model's sensitivity to the position of relevant content (i.e., position bias), we introduce a simple and intuitive metric called the \textbf{Position Sensitivity Index (PSI)}. 
This metric captures the model’s worst-case performance degradation across different positional buckets.
Given a set of position-specific evaluation scores $\mathbf{s} = \{s_1,\ldots,s_k\}$ (e.g., NDCG@10 for each position group), we define PSI as:
\begin{equation}
\label{eq:psi}
\mathrm{PSI}
  \;=\;
  1 - \frac{\min(\mathbf{s})}{\max(\mathbf{s})},
  \quad
  \text{where } \max(\mathbf{s}) > 0.
\end{equation}
Intuitively, PSI measures the relative drop from the best-performing position to the worst-performing one. 
A lower PSI suggests that the model's performance is more consistent across positional buckets, indicating reduced sensitivity to the location of relevant content within the passage.
For example, if the scores are identical across all positions, we have $\min = \max$, resulting in $\mathrm{PSI} = 0$, which signifies complete positional robustness. Conversely, a large gap between $\min$ and $\max$ pushes PSI closer to 1, signaling strong position bias.
Compared to alternative dispersion metrics such as standard deviation or the coefficient of variation (CV)~\cite{coefficient_variation}, PSI provides a worst-case perspective by explicitly quantifying the maximum relative degradation across positions. 
Note that the PSI formulation (Equation~\ref{eq:psi}) is \emph{scale-invariant} in that it captures only the relative variation across positions, independent of the absolute retrieval quality. However, this also means that PSI alone does not reflect a model’s effectiveness. For instance, a model with uniformly low scores (e.g., all NDCG@10 values at 0.1) will have $\mathrm{PSI}=0$ despite being practically ineffective. Therefore, PSI should always be interpreted in conjunction with a measure of overall quality, such as the mean NDCG score across positions, to ensure that both robustness and retrieval performance are properly assessed.

\section{Experiments}
\subsection{Experimental Setup}
We perform a comprehensive evaluation across the full IR pipeline to assess the extent and impact of position bias, covering four distinct categories of retrieval models.

\begin{itemize}[leftmargin=*, noitemsep]
    \item \textbf{Sparse Retrievers}: \texttt{BM25}~\cite{bm25}
    \item \textbf{Dense Retrievers}: \texttt{bge-m3-dense}\footnote{\texttt{bge-m3-dense} denotes the dense retrieval mode of the \texttt{bge-m3} model, where a single vector is generated per query or passage.}~\cite{bge-m3}, \texttt{stella\_en\_400M\_v5}~\cite{stella}, \texttt{text-embedding-3-large}~\cite{text-embedding-3-large}, \texttt{voyage-3-large}~\cite{voyage-3-large}, \texttt{jina-embeddings-v4}~\cite{jinav4}, \texttt{gte-Qwen2-7B-instruct}~\cite{gte-embedding}, \texttt{NV-embed-v2}~\cite{lee2025nvembed}, \texttt{Qwen3-Embedding-0.6B/4B/8B}~\cite{zhang2025qwen3embeddingadvancingtext}
    \item \textbf{ColBERT-style Late-interaction Models}: \texttt{colbertv2.0}~\cite{colbertv2}, \texttt{bge-m3-colbert}\footnote{\texttt{bge-m3-colbert} refers to the late interaction mode of the \texttt{bge-m3} model, where multiple token-level embeddings are generated for each input to enable ColBERT-style retrieval.}~\cite{bge-m3}, \texttt{jina-colbert-v2}~\cite{jina-colbert}
    \item \textbf{Full-interaction Reranker Models}: \texttt{bge-reranker-v2-m3}~\cite{bge-m3}, \texttt{gte-multilingual-reranker-base}~\cite{gte-reranker}, \texttt{bge-reranker-v2-gemma}~\cite{bge-reranker}, \texttt{Qwen3-Reranker-0.6B/4B/8B}~\cite{zhang2025qwen3embeddingadvancingtext}
\end{itemize}

We adopt NDCG@10 as our primary evaluation metric, which captures both retrieval accuracy and ranking quality within the top-10 retrieved results. 
To further quantify worst-case performance variations with respect to the position of relevant content, we introduce the Position Sensitivity Index (PSI) (see Section~\ref{sec:psi}) as a complementary diagnostic metric.
BM25 and dense embedding models are evaluated on the full datasets, whereas the more computationally intensive ColBERT-style and reranker models are assessed on the tiny subsets.  
Experimental results are presented in Table~\ref{tab:retrieval-results}, followed by an in-depth analysis.\footnote{Due to its maximum sequence length of 512 tokens, \texttt{colbertv2.0} is incompatible with the longer-passage setting of \textsc{FineWeb-PosQ}, and is therefore excluded from evaluation on this dataset.}

\subsection{Experimental Results}
\subsubsection{BM25: Naturally Position-Robust}
BM25, a classical sparse retrieval method based on term-matching, exhibits strong robustness to position bias across both \textsc{SQuAD-PosQ} and \textsc{FineWeb-PosQ}.
Its NDCG@10 scores remain relatively stable across all positional buckets, with low PSI values of 0.059 and 0.027, respectively.
This aligns with expectations: BM25 does not encode word order or any positional information, relying solely on keyword overlap.
While this limits its ability to capture deeper semantic relationships, such position-agnostic behavior proves advantageous in scenarios where relevant content appears later in the passage.
BM25 thus serves as a robustness baseline, demonstrating that retrieval quality need not necessarily deteriorate with content position.

\subsubsection{Embedding Models: Widespread Bias}
A wide range of dense embedding-based retrievers, including both open-source models (e.g., \texttt{bge-m3-dense}) and commercial offerings (e.g., \texttt{text-embedding-3-large}), exhibit substantial performance degradation as relevant content appears later in the passage. 
These results align with the head-position bias observed in prior work~\cite{coelho-etal-2024-dwell,CollapseEmbedding}.
Interestingly, the persistence of position bias appears unrelated to model size: from \texttt{Qwen3-Embedding-0.6B} to \texttt{Qwen3-Embedding-8B}, PSI remains consistently high despite increasing model capacity.
Notably, \texttt{voyage-3-large} shows a much higher PSI on \textsc{FineWeb-PosQ} (0.101) than on \textsc{SQuAD-PosQ} (0.036), suggesting potential evaluation leakage in widely used datasets like SQuAD, and underscoring the diagnostic value of the newly constructed \textsc{FineWeb-PosQ} benchmark in revealing latent position bias.
An unexpected case is \texttt{NV-embed-v2}, which displays a reversed trend on \textsc{FineWeb-PosQ}: its lowest NDCG@10 score occurs at the beginning of passages. 
We leave the investigation of this reversal to future work, as it may be attributed to specific architectural design or distributional characteristics of the training corpus.

\subsubsection{ColBERT-style Models: Persistent Bias}
ColBERT-style late-interaction models balance retrieval efficiency and effectiveness by independently encoding queries and passages into multi-vector representations, followed by token-level interactions at inference time.
Although they sometimes outperform dense retrievers in absolute NDCG@10, they still exhibit considerable position bias, especially on longer passages.
For example, \texttt{jina-colbert-v2} suffers a sharp performance drop on \textsc{FineWeb-PosQ}, from 91.69 (beginning) to just 45.91 (end), resulting in a PSI of 0.499—among the highest in our evaluation.
This suggests that late interaction alone cannot fully compensate for position bias introduced during early-stage encoding.
However, variation within the ColBERT family is noteworthy: \texttt{bge-m3-colbert} shows a much lower PSI than \texttt{jina-colbert-v2} on both datasets.
Interestingly, under the same base encoder and training data, \texttt{bge-m3-colbert} clearly outperforms its dense counterpart, \texttt{bge-m3-dense}.
This supports the idea that ColBERT-style training may help mitigate position bias, though it does not fully eliminate it.

\subsubsection{Reranker Models: Effective Mitigation}

Full-interaction reranker models, which apply deep cross-attention between query and passage, demonstrate the highest resilience to position bias among all model classes evaluated. 
All reranker models maintain consistently high NDCG@10 scores across positional buckets, with PSI values uniformly below 0.03. 
For instance, \texttt{bge-reranker-v2-m3} achieves NDCG@10 scores ranging from 93.53 to 94.69 on \textsc{SQuAD-PosQ} (PSI 0.012), and from 94.25 to 96.10 on \textsc{FineWeb-PosQ} (PSI 0.019), indicating a high degree of robustness to the position of relevant content.
These results underscore the strength of full cross-attention, which enables the model to flexibly attend to relevant spans regardless of position.
From a system design perspective, these findings highlight that although dense embedding-based and ColBERT-style retrievers are vulnerable to head-position bias, incorporating an interaction-based reranking stage can substantially mitigate it.
In high-stakes retrieval settings such as RAG applications, integrating a reranker serves as a crucial safeguard, ensuring that relevant information is accurately recognized and appropriately prioritized in the final ranking.
However, this effectiveness relies on the assumption that relevant passages appear in the Top-K retrieval pool, underscoring the importance of the choice of K in practical deployments.

\section{Conclusion}
We conduct a comprehensive study of position bias in the modern IR pipeline.
To enable realistic evaluation, we introduce two position-aware retrieval benchmarks: \textsc{SQuAD-PosQ} and \textsc{FineWeb-PosQ}, repurposed from existing datasets while preserving semantic integrity.
We further propose the Position Sensitivity Index (PSI), a simple and intuitive metric for quantifying position bias across retrieval models.
Our findings reveal that while position bias primarily arises in embedding-based retrievers, it can be substantially mitigated by downstream interaction-based reranker models.

\section*{Limitations}
This work has several limitations that open avenues for future research.
First, our study focuses exclusively on position bias in English text retrieval, and the findings may not directly generalize to multi-lingual or cross-lingual retrieval settings. 
Understanding how position bias manifests in such settings is an important next step.
To this end, we are constructing a highly fine-grained and comprehensive position-aware retrieval benchmark, named \textsc{PosIR}\footnote{\url{https://huggingface.co/datasets/infgrad/PosIR-Benchmark-v1}}, which spans multiple domains and languages, aiming to lay a solid foundation for future research on position bias.
Second, our analysis does not yet provide a theoretical account of why embedding-based retrievers exhibit uneven information distribution in their vector representations. Without such a mechanistic understanding, it is difficult to design principled methods for mitigating position bias. Future work will explore connections to representation theory with the aim of developing more robust and unbiased text representation learning methods.
Finally, our study abstracts away from user interaction effects. In realistic scenarios where multiple relevant passages exist, position bias may interact with human reading or clicking behavior: users tend to notice and rely on information presented earlier, while equally valid content appearing later may be overlooked. Investigating this interaction would yield a more comprehensive understanding of the practical implications of position bias in retrieval systems.

\bibliography{custom}


\appendix

\begin{table*}
\centering
\begin{tabular}{@{}lcccc@{}}
\toprule
& \textbf{SQuAD-PosQ} & \textbf{SQuAD-PosQ-Tiny} & \textbf{FineWeb-PosQ} & \textbf{FineWeb-PosQ-Tiny} \\
\midrule
\# Query & 92,749 & 10,000 & 25,775 & 3,000 \\
Mean Query Length & 10.09 & 10.08 & 13.98 & 14.05 \\
Std Query Length & 3.56 & 3.56 & 4.01 & 4.11 \\
\midrule
\# Passage & 20,233 & -- & 13,902 & -- \\
Min Passage Length & 20 & -- & 500 & -- \\
Max Passage Length & 653 & -- & 1,023 & -- \\
Mean Passage Length & 117.19 & -- & 710.79 & -- \\
Std Passage Length & 50.22 & -- & 132.34 & -- \\
\midrule
\multicolumn{5}{@{}l}{\textbf{Positional Bucket}} \\
0+: [0--100] & 21,220 & 2,252 & \multirow{2}{*}{beginning: 8,467} & \multirow{2}{*}{beginning: 1,000} \\
100+: [100--200] & 16,527 & 1,813 &  &  \\
200+: [200--300] & 13,667 & 1,444 &  \multirow{2}{*}{middle: 8,213} & \multirow{2}{*}{middle: 1,000} \\
300+: [300--400] & 11,514 & 1,210 & &  \\
400+: [400--500] & 10,089 & 1,108 &  \multirow{2}{*}{end: 9,095} & \multirow{2}{*}{end: 1,000} \\ 
500+: [500--3120] & 20,384 & 2,237 &  &  \\
\bottomrule
\end{tabular}
\caption{Statistics of the \textsc{SQuAD-PosQ} and \textsc{FineWeb-PosQ} datasets. The two datasets use different bucketing schemes due to differences in construction methodology.}
\label{table:data_stas}
\end{table*}

\begin{table*}[!h]
\centering
\begin{tabularx}{\textwidth}{c >{\raggedright\arraybackslash}X c}
\toprule
\textbf{No.} & \textbf{Question} & \textbf{Position Tag} \\
\midrule
1& What is the purpose of the computerized vest developed by researchers at Georgia Tech? & beginning \\
2& What doctrine did John Wycliffe dispute, antagonizing the orthodox Church? & middle \\
3& What was the date of George IV's coronation? & end \\
\bottomrule
\end{tabularx}
\caption{Examples from the \textsc{FineWeb-PosQ} dataset with corresponding position tag.}
\label{tab:fineweb_posq_examples}
\end{table*}

\section{Dataset Cards}
\label{sec:data_info}
Table~\ref{table:data_stas} presents summary statistics for the \textsc{SQuAD-PosQ} and \textsc{FineWeb-PosQ} datasets. Note that the two datasets differ in design: SQuAD-PosQ provides a fine-grained character-level positional analysis, while FineWeb-PosQ is constructed with a coarse-grained chunk-based segmentation. These differing granularities serve complementary purposes in analyzing positional effects across diverse settings. Some examples of \textsc{FineWeb-PosQ} are shown in Table~\ref{tab:fineweb_posq_examples}.

\subsection{Distribution Analysis of Answer Positions in SQuAD v2}

\begin{figure}[!h]
 \centering
 \includegraphics[width=1\columnwidth]{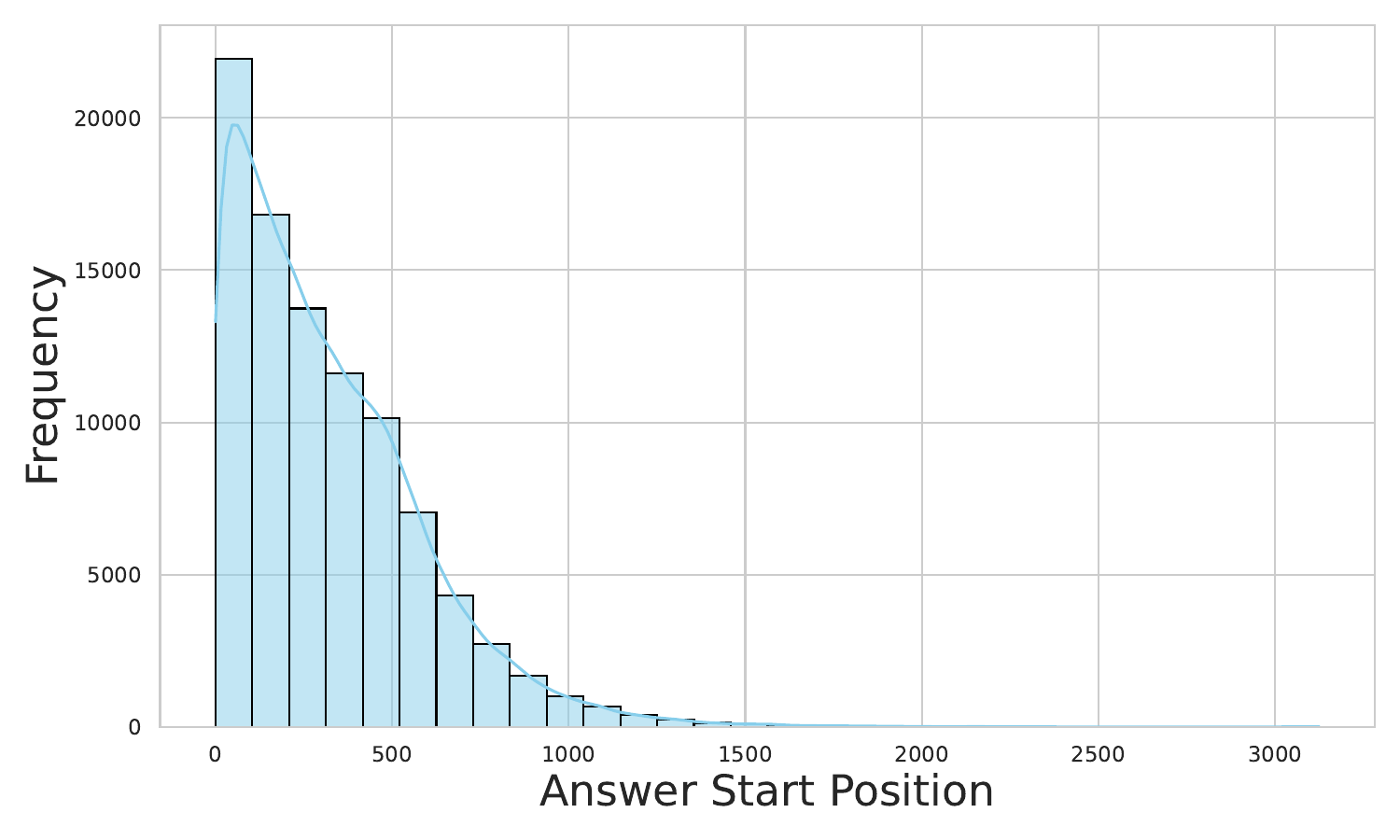} 
 \caption{Distribution of answer start positions in SQuAD v2.}
 \label{fig:squad_dist}
\end{figure}

Figure~\ref{fig:squad_dist} shows the distribution of answer start positions in SQuAD v2, which exhibits a pronounced long-tail pattern: answers tend to appear near the beginning of passages, though a non-negligible portion also occurs in later positions.  
This natural skew makes SQuAD v2 particularly well-suited for analyzing positional effects in retrieval models.

\subsection{\textsc{FineWeb-PosQ} Construction Details}
\label{sec:appendix_fineweb_construction}

\textsc{FineWeb-PosQ} is built from the FineWeb-edu corpus, with the goal of creating a position-aware retrieval benchmark grounded in long, high-quality educational passages. We begin by selecting 13,902 passages between 500 and 1,024 words to ensure sufficient content length. Each passage is globally summarized using \texttt{gpt-4o-mini}, and then split into 256-word chunks using the \texttt{RecursiveCharacterTextSplitter}\footnote{\url{https://python.langchain.com/docs/how_to/recursive_text_splitter/}}.
Each chunk, along with the global summary, is used to generate a (\textit{question}, \textit{answer}, \textit{question\_type}) triplet using \texttt{gpt-4o-mini}. Initially, we experimented with generating only questions, but found that approximately 40\% of them were either unanswerable or misaligned with the source content. To address this, we adopt joint question–answer generation, which substantially improved the quality, answerability, and relevance of the questions produced.
To encourage diversity, we tag each question with a complexity label (\textit{simple} or \textit{complicated}), but all valid samples are retained regardless of complexity type.
Each passage is divided into three equal-length regions: \texttt{beginning}, \texttt{middle}, and \texttt{end}. Each chunk is assigned to a span using a simple rule (Algorithm~\ref{alg:classify_span}). In ambiguous cases where a chunk overlaps with two spans, we assign them to \texttt{middle} span for consistency.

To ensure the dataset is both positionally accurate and high-quality, we implement a two-stage filtering pipeline:

\paragraph{Stage 1: Validating Positional Relevance.}
The initial generation yields 265,865 question–chunk pairs across the 13,902 passages, with 15,961 from the beginning span, 199,742 from the middle span, and 50,162 from the end span.
To ensure each question truly pertains to its labeled position span, we apply a consistency check using three reranker models, including  \texttt{bce-reranker-base\_v1}, \texttt{mmarco-mMiniLMv2-L12-H384-v1}, and \texttt{jina-reranker-v1-turbo-en}. Each model scores the question against the three segments (\textit{beginning}, \textit{middle}, \textit{end}) of its corresponding passage.
Only questions for which all three models agree that the labeled span yields the highest relevance score are retained. This aggressive filtering reduces the dataset to 117,008 questions—13,061 from the beginning span, 65,941 from the middle span, and 38,006 from the end span—ensuring the evaluation set is both positionally precise and semantically reliable.
To further balance the dataset, we downsample each category to 13,061 questions—the size of the smallest category (beginning span)—yielding a total of 39,183 samples.
Then, we verify fine-grained alignment using \texttt{DeepSeek-V3-0324}, prompting it to assign a relevance score (0–4) between the question and each segment (see prompt in Appendix~\ref{pe:rel}). A question is retained only if: (1) The score for its labeled span is 3 or 4. (2) Its score is at least one level higher than any other span.
This LLM-based filtering step takes 36 hours and results in 26,356 questions, with 8,658 from the beginning span, 8,433 from the middle span, and 9,265 from the end span.

\paragraph{Stage 2: Minimizing False Negatives.}
To reduce false negatives (relevant passages incorrectly regarded as irrelevant), we follow the three-step approach from \citet{AIRBench}.
\textit{(1) Recall with Embedding Models.}  
For each question $q_i$, we use \texttt{jina-embedding-v3} to retrieve the top-1,000 relevant passages from the corpus (denoted $L_{\text{recall}} = \{p_1, \dots, p_{1000}\}$). \textit{(2) Pre-label with Rerankers.}  
We rerank $L_{\text{recall}}$ using three rerankers:
\texttt{jina-reranker-v1-turbo-en},
\texttt{bge-reranker-v2-minicpm-layerwise}, and
\texttt{gte-reranker-modernbert-base}.
A passage $p_j$ is labeled positive by model $M$ if its normalized score $r_j(M)$ $\ge$ 0.5.
If a majority of the three models label $d_j$ as positive, we pre-label it as positive; otherwise, negative. 
This step identifies 854 potential false negatives.
\textit{(3) Label with LLMs.}  
We further verify these potential false negatives using three LLMs:
\texttt{deepseek-chat},
\texttt{gemini-2.5-flash},
and \texttt{gpt-4.1-mini}.
Each LLM scores passage relevance from 0–4 (consistent with stage 1). 
A passage is retained as false negative only if at least two LLMs assign a score $\ge$ 3. This confirms 661 high-confidence false negatives.
Given that these high-confidence items affect fewer than 3\% of questions, we remove all associated questions to ensure data purity. 
We do not relabel passages to avoid introducing ambiguity in downstream evaluation.

After the above two filtering stages, the final dataset contains:
\begin{itemize}[nosep]
    \item \textbf{Questions:} 25,775
    \item \textbf{Passages:} 13,902
    \item \textbf{Position Distribution:}
        \begin{itemize}[nosep]
            \item Beginning: 8,467
            \item Middle: 8,213
            \item End: 9,095
        \end{itemize}
\end{itemize}

\begin{algorithm}[H]
\caption{\textsc{Position Tagging}}
\label{alg:classify_span}
\begin{algorithmic}[1]
\Require Total length $z$, chunk start index $m$, end index $n$
\Ensure Return tag: \texttt{beginning}, \texttt{middle}, \texttt{end}
\State $third \gets \lfloor z / 3 \rfloor$
\If{$n < third$}
    \State \Return \{ \texttt{beginning} \}
\ElsIf{$m \ge 2 \cdot third$}
    \State \Return \{ \texttt{end} \}
\Else
    \State \Return \{ \texttt{middle}\}
\EndIf
\end{algorithmic}
\end{algorithm}

\begin{figure}[h]
 \centering
 \includegraphics[width=1\columnwidth]{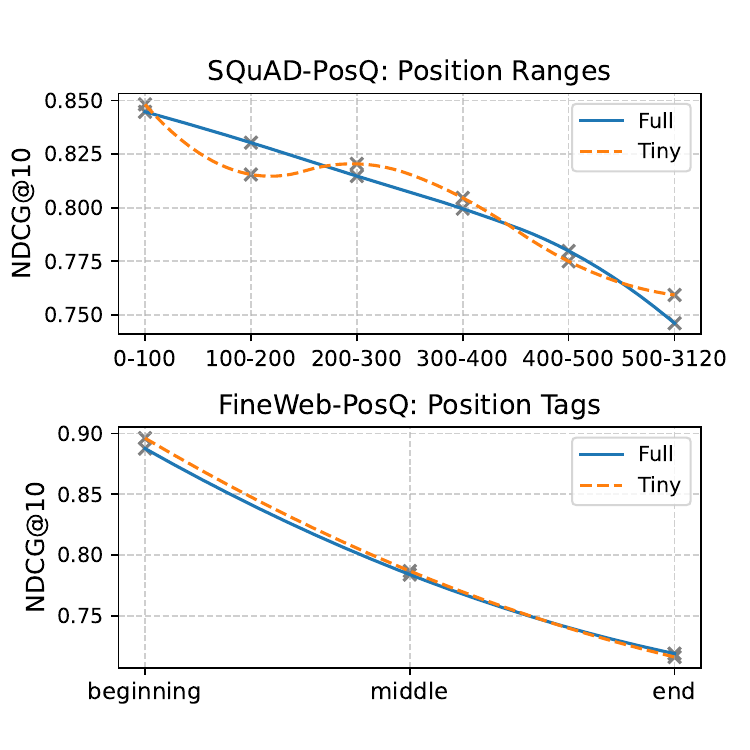} 
 \caption{NDCG@10 scores of \texttt{bge-m3-dense} on Full vs. Tiny Datasets.}
 \label{fig:ndcg_comparison}
\end{figure}

\subsection{Validity of the Sampled Subset}
To empirically verify the validity of the sampled dataset (i.e., \textsc{SQuAD-PosQ-Tiny} and \textsc{FineWeb-PosQ-Tiny}), we conduct preliminary experiments using \texttt{bge-m3-dense} on both the full and tiny versions of each dataset.
Figure~\ref{fig:ndcg_comparison} shows that \texttt{bge-m3-dense} achieves highly consistent NDCG@10 performance between the full and sampled datasets, particularly for \textsc{FineWeb-PosQ}.
These results confirm the feasibility of using the sampled subset to accelerate evaluation for computationally intensive models.
Additionally, the experiments reveal a pronounced head-bias in \texttt{bge-m3-dense}, indicating a tendency to overly prioritize the beginning context while neglecting the middle and end segments during retrieval.

\begin{table*}
  \centering
  \begin{tabular}{clccc}
    \toprule
    \textbf{Dataset} & \textbf{Embedding Model} & \textbf{Full \& Begin} & \textbf{Full \& Middle} & \textbf{Full \& End} \\
    \toprule
     SQuAD v2  & bge-m3-dense  & 0.8777 & 0.7957 & 0.7727 \\
 & stella\_en\_400M\_v5 & 0.8851 & 0.8188 & 0.7930 \\
 & text-embedding-3-large  & 0.8695 & 0.7451 & 0.7251 \\
& voyage-3-large  & 0.8695 & 0.8446 & 0.8335 \\
 & gte-Qwen2-7B-instruct  & 0.8440 & 0.7831 & 0.7456 \\
 & NV-Embed-v2 & 0.7760 & 0.7058 & 0.6854 \\
\midrule
     FineWeb-Edu   & bge-m3-dense  & 0.9201 & 0.8101 & 0.7835 \\
    & stella\_en\_400M\_v5 & 0.9255 & 0.8514 & 0.8280 \\
 & text-embedding-3-large  & 0.8977 & 0.7444 & 0.7805 \\
  & voyage-3-large  & 0.9278 & 0.8837 & 0.8712 \\
 & gte-Qwen2-7B-instruct  & 0.8683 & 0.7775 & 0.7821 \\
 & NV-Embed-v2 & 0.8430 & 0.7402 & 0.7651 \\
    \bottomrule
  \end{tabular}
    \caption{\label{cs_exp}
Cosine similarity between full-text embeddings and segment-level embeddings (beginning, middle, end) across models and datasets. Higher values indicate stronger alignment between the segment and the full-text representation.}
\end{table*}

\subsection{Representation Behavior}
Following the approach of \citet{coelho-etal-2024-dwell}, we compute the cosine similarity between the \textit{full-text} embedding and the embeddings of the \textit{beginning}, \textit{middle}, and \textit{end} segments to examine how embedding models represent different parts of the text.
We selected a random subset of 10,000 passages from the SQuAD v2 dataset (with lengths ranging from 100 to 512 words, average 146 words) and 10,000 passages from the FineWeb-Edu dataset (with lengths ranging from 200 to 500 words, average 339 words).  
As shown in Table~\ref{cs_exp}, we observe that the similarity between the beginning segment and the full text is consistently the highest across most models.  
This suggests that although these models are designed to encode the entire input, they tend to overemphasize its initial portion.  
In contrast, similarity scores for the middle and end segments show a noticeable decline.  
For instance, in \texttt{text-embedding-3-large}, the similarity drops from 0.8695 (full \& beginning) to 0.7451 (full \& middle), and further to 0.7251 (full \& end).  
This tendency is consistent across all models, reinforcing the observation that embedding models exhibit a strong position bias—favoring the beginning of the input while underrepresenting its later parts.

\onecolumn
\section{Prompts}
\label{sec:prompts}
\subsection{Prompt for Summarization}
\begin{tcolorbox}
<task> \\ Given a passage, please paraphrase it concisely. \\ </task> \\

<requirements> \\
- The paraphrase should be concise but not missing any key information. \\
- Please decide the number of words for the paraphrase based on the length and content of the passage, but do not exceed 400 words. \\
- You MUST only output the paraphrase, and do not output anything else. \\
</requirements>  \\

<passage> \code{\{TEXT\}} </passage>
\end{tcolorbox}

\subsection{Prompt for Question Generation}
\begin{tcolorbox}
<task> \\ Given a summary and a chunk of passage, please brainstorm some FAQs for this chunk. \\ </task> \\

<requirements> \\
- The generated questions should be high-frequency and commonly asked by people. \\
- Two types of questions should be generated: simple (e.g., factual questions) and complicated (questions that require reasoning and deep thinking to answer). \\ 
- The majority of the questions you generate should be complicated. \\  
- The answers to the questions must be based on the chunk and should not be fabricated. \\
- You MUST only output the FAQs, and do not output anything else. \\
Note: The FAQ you generate must be based on this chunk rather than the summary!!! The summary is only used to assist you in understanding the chunk. \\
</requirements> \\

<summary> \code{\{SUMMARY\}} </summary> \\

<chunk> \code{\{CHUNK\}} </chunk> \\

Your output should be a JSON List: 
\begin{verbatim}
[
  {
    "question": "Genrated question",
    "answer": "The answer of question",
    "type": "simple or complicated"
  },
  ...
]
\end{verbatim}
\end{tcolorbox}

\subsection{Prompt for Relevance Estimation}
\label{pe:rel}
\begin{tcolorbox}
Evaluate the relevance between the provided query and passage on a scale of 0-4, where:\\
0 = Completely irrelevant\\
1 = Slightly relevant (minimal connection)\\
2 = Moderately relevant (partial match)\\
3 = Highly relevant (covers most aspects)\\
4 = Perfectly relevant (document fully addresses query)\\
\\
Query: \{query\}\\
\\
Passage: \{passage\}\\
\\
Output only a single integer from {0,1,2,3,4} without any additional text, explanations, or formatting. Higher values indicate stronger relevance.
\end{tcolorbox}

\end{document}